# The ARM Scalable Vector Extension

Nigel Stephens, Stuart Biles, Matthias Boettcher, Jacob Eapen, Mbou Eyole, Giacomo Gabrielli,
Matt Horsnell, Grigorios Magklis, Alejandro Martinez, Nathanael Premillieu, Alastair Reid,
Alejandro Rico, Paul Walker
ARM

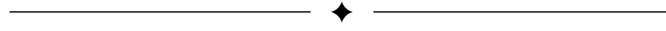

In this paper we describe the ARM Scalable Vector Extension (SVE). Several goals guided the design of the architecture. First was the need to extend the vector processing capability associated with the ARM AArch64 execution state to better address the compute requirements in domains such as high performance computing (HPC), data analytics, computer vision and machine learning. Second was the desire to introduce an extension that can scale across multiple implementations, both now and into the future, allowing CPU designers to choose the vector length most suitable for their power, performance and area targets. Finally, the architecture should avoid imposing a software development cost as the vector length changes and where possible reduce it by improving the reach of compiler auto-vectorization technologies.

We believe SVE achieves these goals. It allows implementations to choose a vector register length between 128 and 2048 bits. It supports a vector length agnostic programming model which allows code to run and scale automatically across all vector lengths without recompilation. Finally, it introduces several innovative features that begin to overcome some of the traditional barriers to auto-vectorization.

## 1 INTRODUCTION

Architecture extensions are often somewhat conservative when they are first introduced and are then expanded as their potential becomes better understood and transistor budgets increase. Over the course of 15 years, ARM has expanded and improved support for SIMD. Starting with 32-bit integer-only SIMD instructions using the integer register file in ARMv6-A [1] to 64 and 128-bit SIMD instructions sharing the floating point register file in the most recent incarnation of Advanced SIMD in ARMv7-A and ARMv8-A [2].

These extensions efficiently target media and image processing workloads, which typically process structured data using well-conditioned DSP algorithms. However, as our partners continue to deploy ARMv8-A into new markets we have seen an increasing demand for more radical changes to the ARM SIMD architecture, including the introduction of well known technologies such as gather-load and scatter-store, per-lane predication and longer vectors.

But this raises the question, what should that vector length be? The conclusion from over a decade of research into vector processing, both within ARM [3], [4], and taking inspiration from more traditional vector architectures, such as the CRAY-1 [5], is that there is no single preferred vector length. For this reason, SVE leaves the vector length as an implementation choice (from 128 to 2048 bits, in increments of 128 bits). Importantly the programming model adjusts dynamically to the available vector length, with no need to recompile high-level languages or to rewrite hand-coded SVE assembly or compiler intrinsics.

Of course, longer vectors are only part of the solution and achieving significant speedup also requires high vector utilization. At a high level, the key SVE features enabling improved auto-vectorization support are:

- **Scalable vector length** increasing parallelism while allowing implementation choice.
- **Rich addressing modes** enabling non-linear data accesses.
- **Per-lane predication** allowing vectorization of loops containing complex control flow.
- **Predicate-driven loop control and management** reduces vectorization overhead relative to scalar code.
- **A rich set of horizontal operations** applicable to more types of reducible loop-carried dependencies.
- **Vector partitioning and software-managed speculation** enabling vectorization of loops with data-dependent exits.
- **Scalarized intra-vector sub-loops** permitting vectorization of loops with more complex loop-carried dependencies.

The remainder of this paper is organized as follows. In Section 2 we introduce the SVE architecture and the vector length agnostic programming model. Saliant features of the architecture are demonstrated with examples. Section 3 discusses the implications for compilers with a focus on auto-vectorization. Section 4 outlines some of the implementation challenges. Section 5 presents early performance data based on a representative model and an experimental compiler. Finally, Section 6 concludes.

## 2 SVE OVERVIEW

In this section we more fully describe the architectural state and key features introduced by SVE and, where appropriate,

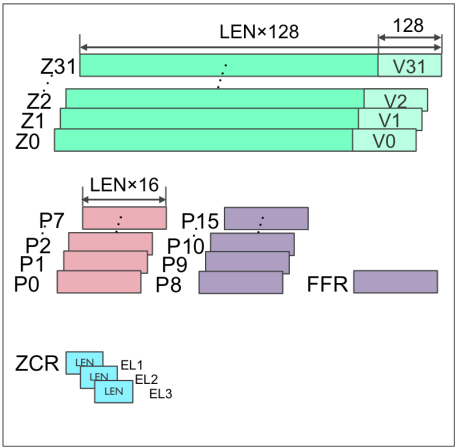
(a) SVE registers

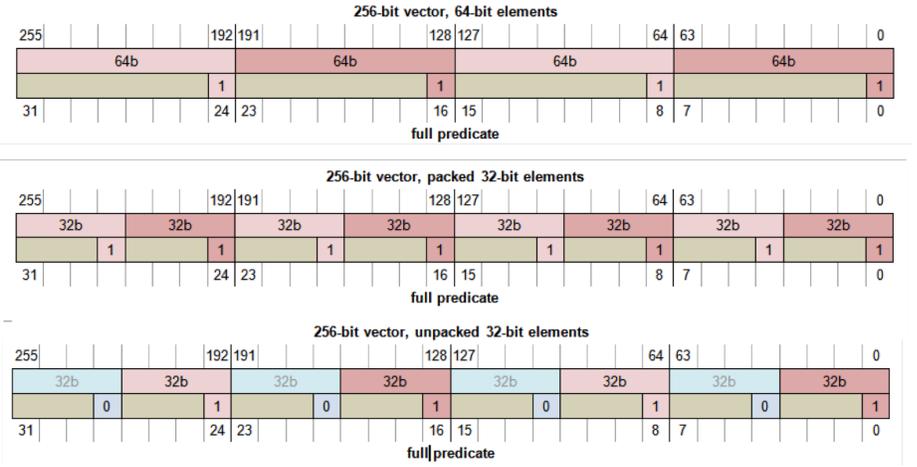
(b) SVE predicate organization

Fig. 1: SVE architectural state - vector registers (Z0–Z31), predicate registers (P0–P15), the first-fault register (FFR) and exception-level specific control registers (ZCR_EL1–ZCR_EL3)

we illustrate these with code examples.

### 2.1 Architectural State

SVE introduces new architectural state, shown in Fig. 1a. This state provides thirty-two new scalable vector registers (Z0–Z31). Their width is implementation dependent within the aforementioned range. The new registers extend the thirty-two 128-bit wide Advanced SIMD registers (V0–V31) to provide scalable containers for 64-, 32-, 16-, and 8-bit data elements.

Alongside the scalable vector registers, are sixteen scalable predicate registers (P0–P15) and a special purpose first-faulting register (FFR).

Finally, a set of control registers (ZCR_EL1–ZCR_EL3) are introduced which give each privilege level the ability to virtualize (by reduction) the effective vector width.

### 2.2 Scalable Vector Length

Within a fixed 32-bit encoding space, it is not viable to create a different instruction set every time a different vector width is demanded. SVE radically departs from this approach in being vector length agnostic, allowing each implementation to choose a vector length that is any multiple of 128 bits between 128 and 2048 bits (the current architectural upper limit). Not having a fixed vector length allows SVE to address multiple markets with implementations targeting different performance-power-area optimization points.

This novel aspect of SVE enables software to scale gracefully to different vector lengths without the need for additional instruction encodings, recompilation or software porting effort. SVE provides the capabilities for software to be vector length agnostic through the use of *vector partitioning* while also supporting more conventional SIMD coding styles requiring fixed-length, multiple-of-N or power-of-two sub-vectors.

### 2.3 Predicate-centric Approach

Predication is central to the design of SVE. In this section we describe the predicate register file, its interaction with other

```c
void daxpy(double *x, double *y, double a, int n)
{
    for (int i = 0; i < n; i++) {
        y[i] = a*x[i] + y[i];
    }
}
```
(a) Daxpy C code

```
// x0 = &x[0], x1 = &y[0], x2 = &a, x3 = &n
daxpy_:
  ldrsw x3, [x3]             // x3=*n
  mov   x4, #0               // x4=i=0
  ldr   d0, [x2]             // d0=*a
  b     .latch
.loop:
  ldr   d1, [x0, x4, lsl #3] // d1=x[i]
  ldr   d2, [x1, x4, lsl #3] // d2=y[i]
  fmadd d2, d1, d0, d2       // d2+=x[i]*a
  str   d2, [x1, x4, lsl #3] // y[i]=d2
  add   x4, x4, #1           // i+=1
.latch:
  cmp   x4, x3               // i < n
  b.lt  .loop                // more to do?
  ret
```
(b) Daxpy ARMv8-A scalar code

```
// x0 = &x[0], x1 = &y[0], x2 = &a, x3 = &n
daxpy_:
  ldrsw   x3, [x3]                // x3=*n
  mov     x4, #0                  // x4=i=0
  whilelt p0.d, x4, x3            // p0=while(i++<n)
  ld1rd   z0.d, p0/z, [x2]        // p0:z0=bcast(*a)
.loop:
  ld1d    z1.d, p0/z, [x0, x4, lsl #3] // p0:z1=x[i]
  ld1d    z2.d, p0/z, [x1, x4, lsl #3] // p0:z2=y[i]
  fmla    z2.d, p0/m, z1.d, z0.d  // p0?z2+=x[i]*a
  st1d    z2.d, p0, [x1, x4, lsl #3]   // p0?y[i]=z2
  incd    x4                      // i+=(VL/64)
.latch:
  whilelt p0.d, x4, x3            // p0=while(i++<n)
  b.first .loop                   // more to do?
  ret
```
(c) Daxpy ARMv8-A SVE code

Fig. 2: Equivalent C, scalar and SVE representations of the Daxpy kernel

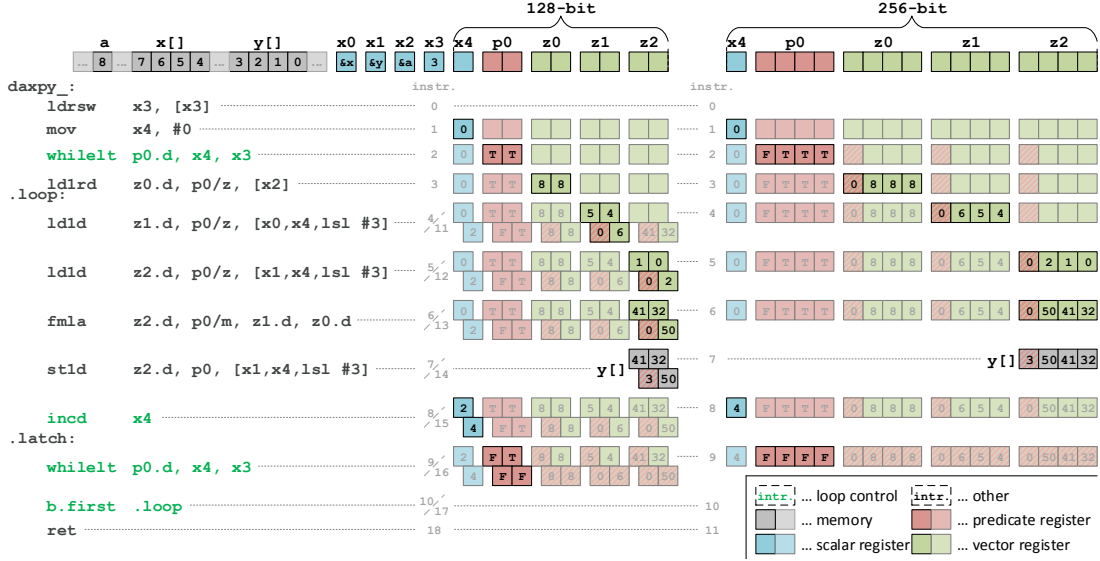

Fig. 3: Cycle by cycle example of daxpy with $n = 3$ and hardware vector lengths of 128- and 256-bit

TABLE 1: SVE condition flags overloading

| Flag | SVE | Condition |
|---|---|---|
| N | First | Set if first element is active |
| Z | None | Set if no element is active |
| C | !Last | Set if last element is not active |
| V | | Scalarized loop state, else zero |

architectural state, and how the use of predicates enables a number of advanced features.

### 2.3.1 Predicate registers

The predicate register file is organized and used as follows:

**Sixteen scalable predicate registers (P0–P15)** Control of general memory and arithmetic operations is restricted to P0–P7, however, predicate-generating instructions (e.g. vector compares), and instructions working solely on predicates (e.g. logical operations) can use the full set P0-P15 (see Fig. 1a). This balance has been validated by analyzing compiled and hand-optimized codes, and mitigates the predicate register pressure observed on other architectures [6].

**Mixed element size control** Each predicate consists of eight enable bits per 64-bit vector element, allowing down to per byte-granularity. For any given element size only the least significant bit is used as the enable. This is important for vectorizing code containing multiple data types (see Fig. 1b).

**Predicate conditions** Predicate generating instructions (e.g. vector compares and logical operations) in SVE reuse the AArch64 *NZCV* condition code flags, which in the context of predication are interpreted differently as shown in TABLE 1.

**Implicit order** Predicates are interpreted in an implicit least- to most-significant element order, corresponding to a an equivalent sequential ordering. We refer to the *first* and *last* predicate elements and associated conditions with respect to this order.

### 2.3.2 Predicate-driven loop control

Predication is used for fundamental loop control in SVE. In other SIMD architectures that support predication such as ICMI [7] and AVX-512 [8], generating the governing predicate of a loop often requires a test of the induction variable. This is typically done by calculating the sequence of incrementing values in a vector register and then using that vector as input to a predicate-generating instruction (e.g. a vector comparison).

There are two sources of overhead associated with this approach. First, a vector register is "wasted" to store the sequence, and second, auto-vectorizing compilers tend to align all SIMD instructions in a loop to the largest element size, thus resulting in a potential loss of throughput when the size of the induction variable is larger than the size of the data elements processed within the loop.

To overcome these limitations in common-case scenarios, SVE includes a family of `while` instructions that work with scalar count and limits to populate a predicate with the loop iteration controls that would have been calculated by the corresponding sequential loop. Note that if the loop counter is close to the maximum integer value, then `while` will handle potential wrap-around behaviour consistently with the semantic of the original sequential code. Similarly to other predicate-generating instructions, `while` also updates the condition flags.

The example in Fig. 2 demonstrates some of these concepts. It shows the Daxpy[1] loop in both C, ARMv8-A scalar assembly, and ARMv8-A SVE assembly[2]. Notice that there is no overhead in instruction count for the SVE version Fig. 2c when compared to the equivalent scalar code Fig. 2b, which allows a compiler to opportunistically vectorize loops with an unknown trip count.

The same example is illustrated in Fig. 3, which steps through the SVE version of the code, at both 128-bit and 256-bit vector lengths. The diagram shows the intermediate architectural state, for predicate P registers and vector Z registers, and the relative instructions required to process four array elements at two different vector lengths. We refer the reader to the SVE reference manual [2], and the VLA programming whitepaper [9] for more guidance on the instructions used, but sadly a full narrative for this example is beyond the word limit of this paper.

---
1. Daxpy - double precision Ax plus y.
2. Note that for brevity these examples are sub-optimal.

### 2.3.3 Fault-tolerant Speculative Vectorization

To vectorize loops with data-dependent termination conditions, software has to perform some operations speculatively before the condition can be resolved. For some types of instructions, such as simple integer arithmetic, this is harmless because there are no side effects. However, for instructions that may have side effects when operating on invalid addresses, it is necessary to have mechanisms in place to avoid those side effects.

In SVE, this is achieved with the introduction of a *first-fault* mechanism for vector load instructions. This mechanism suppresses memory faults if they do not result from the first active element[3] in the vector. Instead, the mechanism updates a predicate value in the first-fault register (FFR) to indicate which elements were not successfully loaded following a memory fault.

Fig. 4 shows an example with a gather load that speculatively loads from addresses held in register Z3. In the first iteration, FFR is initialized to all *true*. The translations of A[0] and A[1] succeed, but the address A[2] is invalid (e.g., unmapped) and it fails without taking a trap. Instead, positions corresponding to A[2] and A[3] in the FFR are set to *false*. For the second iteration, the positions corresponding to A[0] and A[1] in the instruction predicate register P1 will be set to *false* and FFR to *all-true* again. In this case, A[2] fails again but, since it is now the first active element, traps to the OS to service the fault or terminate the program if it is an illegal access.

Fig. 5 shows how speculative vectorization with the first-faulting mechanism allows vectorization of the *strlen* function. The `ldff1b` instruction loads the characters in s and sets the FFR positions starting from the first faulty address to *false*, so only the successful positions remain set to *true*. The FFR predicate value is transferred to P1, which predicates the subsequent instructions that check for the end of string character. Following a memory fault, the next loop iteration will retry the faulty access but now as the first active element and will trap.

This flexible mechanism allows fault-tolerant speculative vectorization of loops with data-dependent terminations, such as *strlen*, that would not be vectorized safely otherwise.

### 2.3.4 Dynamic Exits

The previous section shows an example of how SVE vectorizes a loop without an explicit iteration count. The technique that has been used is called *vector partitioning* and it consists of operating on a partition of safe elements in response to dynamic conditions.

Partitions are implemented using predicate manipulating instructions and are inherited by nested conditions

---

3. Note that we use the term *active element* to the refer to an element within a vector for which the governing predicate for that element is set true.

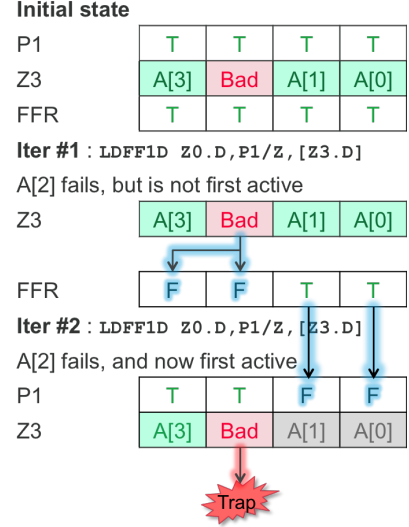

Fig. 4: Example of speculative gather load controlled by the first-fault register FFR

```
1 int strlen(const char *s) {
2     const char *e = s;
3     while (*e) e++;
4     return e - s;
5 }
```

(a) Strlen C code

```
1 // x0 = s
2 strlen:
3   mov x1, x0       // e=s
4 .loop:
5   ldrb x2, [x1], #1 // x2=*e++
6   cbnz x2, .loop   // while(*e)
7 .done:
8   sub x0, x1, x0   // e-s
9   sub x0, x0, #1   // return e-s-1
10  ret
```

(b) Unoptimized ARMv8 scalar strlen

```
1 // x0 = s
2 strlen:
3   mov   x1, x0              // e=s
4   ptrue p0.b                // p0=true
5 .loop:
6   setffr                    // ffr=true
7   ldff1b z0.b, p0/z, [x1]   // p0:z0=ldff(e)
8   rdffr  p1.b, p0/z         // p0:p1=ffr
9   cmpeq  p2.b, p1/z, z0.b, #0 // p1:p2=(*e==0)
10  brkbs  p2.b, p1/z, p2.b   // p1:p2=until(*e==0)
11  incp   x1, p2.b           // e+=popcnt(p2)
12  b.last .loop              // last=>!break
13  sub    x0, x1, x0         // return e-s
14  ret
```

(c) Unoptimized ARMv8 SVE strlen

Fig. 5: Equivalent C, scalar and SVE representations of strlen

and loops. In this way, *vector partitioning* is a natural way to deal with uncounted loops with data-dependent exits (do-while, break, etc). Vectorized code must guarantee that operations with side-effects following a loop exit must not be architecturally performed. This is achieved by operating on a *before-break* vector partition, then exiting the loop if a break was detected.

Fig. 5 shows how *vector partitioning* is used to vectorize the strlen function. Section 2.3.3 describes how the ldff1b instruction loads values speculatively. The rdffr instruction reports the partition of safely loaded values. This is used by the cmpeq instruction to just compare the safe (correct) values with zero. Furthermore, the brkbs instruction generates a sub-partition bounded by the loop break condition and computes the *last* condition accordingly.

### 2.3.5 Scalarized intra-vector sub-loops

Complex loop-carried dependencies are a significant barrier to vectorization that SVE can help to address. One approach to overcoming those is to split a loop (loop fission) into an explicitly serial part, allowing the rest of the loop to be profitably vectorized. However, in many cases, the cost of unpacking and packing the data to work on it serially negates any performance uplift. To reduce that cost, SVE provides support for serially processing elements in place within a vector.

An example of this problem is traversing a linked list, as there is a loop-carried dependency between each iteration (Fig. 6a). By applying loop fission the loop is split into a serial pointer chase followed by a vectorizable loop (Fig. 6b).

Fig. 6c shows how SVE vectorizes this code. The first part is the serialized pointer chase. The pnext instruction allows operating on active elements one-by-one by setting P1 to the next active element and computing the *last* condition. The cpy instruction inserts scalar register X1 into the vector register Z1 at this position. Then, the ctermeq instruction is used to detect the end of the list (*p == NULL*) or the end of the vector (by testing the *last* condition set by the pnext instruction). The b.tcont branch checks this and continues with the serial loop if more pointers are available and needed.

The partition of loaded pointers is computed into P2. In this case, the vectorized loop is just performing the exclusive-or operation. Finally, all the vector elements in *z0* are combined with a horizontal exclusive-or reduction using the eorv instruction. In this example, the performance gained may not be sufficient to justify using vectorization for this loop, but it serves to illustrate the principle applicable to more profitable scenarios.

### 2.4 Horizontal Operations

Another problem for traditional SIMD processing is the presence of dependencies across multiple loop iterations. In many cases, these dependencies can be resolved using a simple horizontal *reduction* operation. Unlike normal SIMD instructions, horizontal operations are a special class of instructions that operate across the elements of the same vector register. SVE has a rich set of horizontal operations including both logical, integer and floating-point reductions

```
struct {uint64 val; struct node *next} *p;
uint64 res = 0;
for (p = &head; p != NULL; p = p->next)
  res ^= p->val;
```

(a) Linked-list loop carried dependency

```
for (p = &head; p != NULL; ) {
  for (i = 0; p != NULL && i < VL/64; p = p->next)
    p'[i++] = p;      // collect up to VL/64 pointers
  for (j = 0; j < i; j++)
    res ^= p'[j]->val; // gather from pointer vector
}
```

(b) Split loop: serial pointer chase, and vectorizable loop

```
 // P0 = current partition mask
  dup    z0.d, #0              // res'= 0
  adr    x1, head              // p = &head
loop:
  // serialized sub-loop under P0
  pfalse p1.d                  // first i
inner:
  pnext  p1.d, p0, p1.d        // next i in P0
  cpy    z1.d, p1/m, x1        // p'[i]=p
  ldr    x1, [x1, #8]          // p=p->next
  ctermeq x1, xzr              // p==NULL?
  b.tcont inner                // !(term|last)
  brka   p2.b, p0/z, p1.b      // P2[0..i] = T
  // vectorized main loop under P2
  ld1d   z2.d, p2/z, [z1.d, #0] // val'=p ->val
  eor    z0.d, p2/m, z0.d, z2.d // res'^=val'
  cbnz   x1, loop              // while p!=NULL
  eorv   d0, p0, z0.d          // d0=eor(res')
  umov   x0, d0                // return d0
  ret
```

(c) Split-loop ARMv8 SVE code

Fig. 6: Separation of loop carried dependencies for partial vectorization of linked-lists

as well as strictly-ordered reduction for floating-point (e.g. fadda[4]).

## 3 COMPILING FOR SVE

The previous sections illustrate many aspects of how to program SVE. Translating these techniques into a compiler required us to rethink our compilation strategy because of the impact of wide vectors, vector length agnosticism, predication and speculative vectorization with first-faulting loads.

### 3.1 Wide Vectors and Vector Length Agnosticism

When compiling for fixed-length vectors such as Advanced SIMD, one approach is "Unroll and Jam" where a loop is first unrolled by the number of elements in a vector and then corresponding operations from each iteration are merged together into vector operations. This approach is clearly incompatible with a scalable vector length because the length is not known at compile time. The solution is for the vectorizer to directly map scalar operations to corresponding vector operations. A second challenge is that knowledge of the constant vector length, *VL*, often plays a critical role in vectorization. For example, when handling induction variables, one might initialize a vector with the numbers $[0, 1, \ldots VL-1]$ by loading it from memory and increment the vector by $VL$ each time round the loop. SVE addresses this with a family of instructions where the current vector length is an implicit operand, for example the `index` instruction initializes a vector induction variable and the `inc` instructions advance an induction variable based on the current vector length and specified element size.

---

4. fadda - is a *strictly-ordered* floating-point add reduction, accumulating into a scalar register.

Vector length agnosticism also impacts register reads and writes to the stack such as spilling/filling due to register pressure or passing a vector argument as part of the function calling convention. Supporting this inside an existing compiler is challenging because compilers normally assume that all objects are at a constant offset in the stack frame and these constants are used in many places within the compiler. Our solution is to introduce stack regions, where each region determines what the constant offset represents. For existing stack regions, constants remain unchanged (i.e., byte offsets) but regions for SVE registers are dynamically allocated and the load and store constant offsets within that region are implicit multiples of $VL$.

### 3.2 Predication

Predicates are introduced by a conventional "if conversion" pass that replaces if-statements with instructions to calculate predicates and then uses the appropriate predicate on each operation dominated by the control of the condition.

This approach is extended to handle conditional branches out of the loop, by inserting a `brk` instruction that generates a vector partition where only those lanes prior to the loop exit condition are active.

### 3.3 Floating Point

Floating point is a challenge for compiler vectorization because vectorizing a reduction loop will change the order of floating point operations which may give a different result from the original scalar code. Programmers then have to choose whether they want consistent results by disabling vectorization or whether they can tolerate some variation to achieve better performance.

Vector length agnosticism introduces more variation because a different vector length could cause a different ordering and, therefore, a different result. SVE mitigates this by providing `fadda` that allows a compiler to vectorize those cases where the precise order of floating point additions is critical to correctness.

### 3.4 Speculative Vectorization

We were able to implement all of the above changes as extensions of LLVM[10] compiler's existing vectorization pass but it was not feasible to support speculative vectorization within the existing pass. We implemented speculative vectorization in a separate pass whose current focus is on expanding loop coverage rather than generating the highest quality code.

The new vectorizer works in largely the same way as LLVM's current vectorizer but has more advanced predicate handling to support loops with multiple exits. It splits the loop body into multiple regions each under the control of a different predicate. Broadly speaking, these regions represent instructions that are safe to always execute, instructions that are required to calculate a conditional exit predicate and those instructions that occur after the conditional exit. For the latter two regions we make use of the first faulting loads and partitioning operations.

## 4 IMPLEMENTATION CHALLENGES

Encoding space is a scarce resource for a fixed-width instruction set so one of the early design constraints for SVE was to limit its overall encoding footprint so as to retain space for future expansion of the A64 instruction set. To achieve that, a few different solutions have been adopted in different areas of the ISA:

**Constructive vs. destructive forms**: While compilers find it desirable to have constructive forms of instructions (i.e. supporting a destination operand distinct from the source operands), the encoding space required to provide both predication and constructivity for the entire set of data-processing operations would have easily exceeded the projected encoding budget (three vector and one predicate register specifier would require nineteen bits alone, without accounting for other control fields). The tradeoff that SVE makes is to provide only destructive predicated forms of most data-processing instructions, while providing constructive unpredicated forms of only the most common opcodes.

**Move Prefix**: To fulfill the need for fully constructive predicated forms, SVE introduces a `movprfx` instruction, which is trivial for hardware to decode and combine with the immediately following instruction to create a single constructive operation. However, it is also permitted to be implemented as a discrete vector copy, and the result of executing the pair of instructions with or without combining them is identical. SVE supports both predicated (zeroing or merging) and unpredicated forms of `movprfx`.

**Restricted access to predicate registers**: To further reduce the encoding space, as mentioned in section 2.3, predicated data-processing instructions are restricted to access predicate registers P0-P7, while predicate-generating instructions can typically access all 16 registers.

The solutions described above keep the encoding space used by SVE below twenty-eight bits (Fig. 7). Aside from encoding space, a key implementation concern for SVE is the extra hardware cost required to support the additional functionality above and beyond Advanced SIMD. A key decision was to overlay the new vector register file on the existing SIMD and floating-point register file (see Fig. 1a), thus minimizing the area overhead, which is especially relevant for smaller cores.

Furthermore the Advanced SIMD and floating-point instructions are required to zero the extended bits of any vector register which they write, avoiding partial updates, which are notoriously hard to handle in high-performance microarchitectures. In addition, the vast majority of SVE operations can be mapped efficiently onto an existing Advanced SIMD datapath and functional units, with the necessary modifications for predication and to support a larger width, if required.

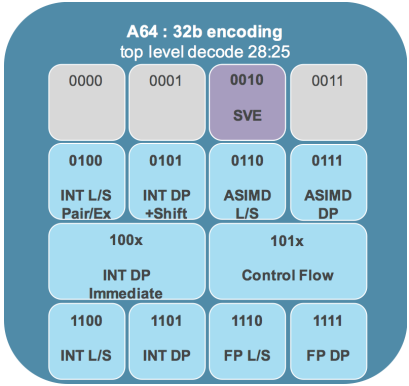

(a) A64 top-level encoding structure, with SVE occupying a single 28-bit region.

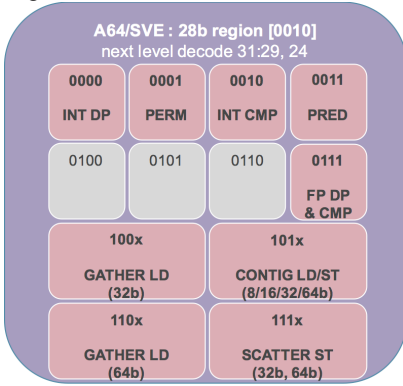

(b) SVE encoding structure. Some room for future expansion is left in this region.

Fig. 7: SVE encoding footprint.

TABLE 2: Model configuration parameters.

| | |
|---|---|
| L1 instruction cache | 64KB, 4-way set-associative, 64B line |
| L1 data cache | 64KB, 4-way set-associative, 64B line, 12 entry MSHR |
| L2 cache | 256KB, 8-way set-associative, 64B line |
| Decode width | 4 instructions/cycle |
| Retire width | 4 instructions/cycle |
| Reorder buffer | 128 entries |
| Integer execution | 2 x 24 entries scheduler (symmetric ALUs) |
| Vector/FP execution | 2 x 24 entries scheduler (symmetric FUs) |
| Load/Store execution | 2 x 24 entries scheduler (2 loads / 1 store) |

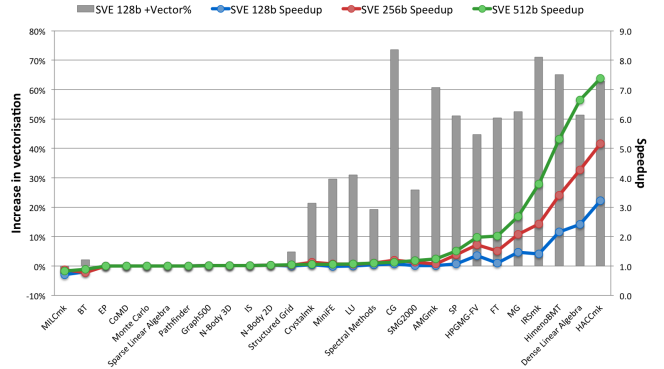

Fig. 8: Performance of SVE at 3 different vector lengths (cf. Section 5)

While a wider datapath for vector processing is a key requirement for exploiting the data-level parallelism exhibited by several workloads in the HPC domain, it has to be coupled with a corresponding improvement in memory access capabilities and bandwidth. SVE offers a wide range of contiguous loads and stores with a rich set of addressing modes, and load-and-broadcast instructions which are able to duplicate a single element across a vector, typically as part of the load/store datapath, thus removing the need for additional permutes in common cases.

Gather-scatter memory operations are an enabling feature that permit vectorization of loops accessing discontiguous data that would otherwise be unprofitable. They can benefit from advanced vector load/store units that can exploit the parallelism [4], but are also amenable to more conservative approaches that crack them into micro operations, so long as this is not noticeably slower than a sequence of scalar loads or stores.

## 5 SVE Performance

We expect the SVE architecture to be implemented by multiple ARM partners, on a variety of micro-architectures. Thus for our evaluation of SVE we have used several representative microarchitecture models. For the results we present here we have chosen a single model of a *typical*, medium sized, out-of-order microprocessor, that does not correspond to any real design, but that we believe gives a fair estimate of what to expect with SVE. The main parameters of this model are shown in Table 2.

Instruction execution and register file access latencies in the model are set to correspond to RTL synthesis results. For operations that cross lanes (i.e. vector permutes and reductions), the model takes a penalty proportional to *VL*. The cache in the model is a true dual-ported cache with the maximum access size being the full cache line, 512 bits. Accesses crossing cache lines take an associated penalty.

Our evaluation uses an experimental compiler, able to auto-vectorize code for SVE. We have chosen a set of high performance computing applications, from various well known benchmark suites [11], [12], [13], [14], [15], [16], [17], [18]. At present our compiler supports only C and C++, so the choice of benchmarks are restricted to these languages. We use the original code from the publicly available versions of the benchmarks with minor modifications in a few cases to help guide the auto-vectorizer (*e.g.*, adding `restrict` qualifiers or OpenMP `simd` pragmas).

Figure 8 shows the results of our evaluation. We are comparing Advanced SIMD with three different configurations of SVE with 128 bit, 256 bit, and 512 bit vector length respectively. All four simulations use the same processor configuration, but vary the vector length. There are two types of results in the figure. The lines in the graph show the speedup of each SVE configuration compared to Advanced SIMD. The bars in the graph show the *extra* vectorization achieved with SVE compared to Advanced SIMD. This is measured as the percentage of dynamically executed vector instructions at a vector length of 128 bits.

One immediate observation from the results is that SVE achieves higher vector utilization than Advanced SIMD.

This is due to all the features we have introduced to the architecture that allow the compiler to vectorize code with complex control flow, non-contiguous memory accesses, etc. For this reason SVE can achieve speedups of up to 3× even when the vectors are the same size as Advanced SIMD. For example, in the particular case of *HACCmk*, the main loop has two conditional assignments that inhibit vectorization for Advanced SIMD, but the code is trivially vectorized for SVE.

The figure also demonstrates the benefits of vector length agnostic code. We can clearly see how performance scales, simply by running the same executable on implementations of SVE with larger vectors. This is one of the greatest benefits of SVE.

There are three clearly identifiable categories of benchmark. On the right of the figure we can see a group of benchmarks that show much higher vectorization with SVE, and performance that scales well with the vector length (up to 7×). Some of these benchmarks do not scale as well as others, and this is mainly due to the use of gather-scatter operations. Although these instructions enable vectorization, our assumed implementation conservatively cracks the operations and so does not scale with vector length. In other cases, such as in *HimenoBMT*, the reason for poor scaling is bad instruction scheduling by the compiler.

On the left of the figure we can see a group of benchmarks for which there is minimal, in some cases zero, vector utilization for both Advanced SIMD and SVE. Our investigations show that this is due to the way the code is structured or limitations of the compiler rather than a shortcoming of the architecture. For example, we know that by restructuring the code in *CoMD* we can achieve significant improvement in vectorization and execution time. Also, it should be noted that the toolchain used for these experiments did not have vectorized versions of some basic math library functions such as `pow()` and `log()`, which inhibit vectorization of loops in some cases, *e.g.*, in *EP*. Finally, there are cases where the algorithm itself is not vectorizable, for example in *Graph500*, where the program mostly traverses graph structures following pointers. We do not expect SVE to help here, unless the algorithm is refactored with vectorization in mind.

The third group of benchmarks includes a few where the compiler has vectorized significantly more code for SVE than for Advanced SIMD, but we do not see much performance uplift. All these cases are due to code generation issues with the compiler which we are currently addressing. In *SMG2000* for example, a combination of bad instruction selection compounded by extensive use of gather loads results in very small benefit for SVE. It is worth noting here that the Advanced SIMD compiler cannot vectorize the code at all.

*MILCmk* is another interesting case, where a series of poor compiler decisions contributes to performance loss for SVE compared to Advanced SIMD. In this case the compiler decides to vectorize the outermost loop in a loop nest generating unnecessary overheads (the Advanced SIMD compiler vectorizes the inner loop), and does not recognize some trivially vectorizable loops as such.

We expect that over time, with improvements to compilers and libraries, many of these issues will be resolved.

# 6 CONCLUSIONS

SVE opens a new chapter for the ARM architecture in terms of the scale and opportunity for increasing levels of vector processing on ARM processor cores. It is early days for SVE tools and software, and it will take time for SVE compilers and the rest of the SVE software ecosystem to mature. HPC is the current focus and catalyst for this compiler work, and creates development momentum in areas such as Linux distributions and optimized libraries for SVE, as well as in tools and software from ARM and third parties.

We are already engaging with key members of the ARM partnership, and are now broadening that engagement across the open-source community and wider ARM ecosystem to support development of SVE and the HPC market, enabling a path to efficient Exascale computing.

# 7 ACKNOWLEDGEMENTS

SVE has been a multi-team, multi-site and multi-year effort across ARM's research and product groups. We would like to take the opportunity to thank everyone involved in the development of SVE, and those who helped review this paper.

# REFERENCES

[1] ARM Ltd, *ARM Architecture Reference Manual (ARMv6 edition)*. ARM Ltd, 2005.
[2] "A-Profile Architecture Specifications." https://developer.arm.com/products/architecture/a-profile/docs.
[3] M. Woh, Y. Lin, S. Seo, S. Mahlke, T. Mudge, C. Chakrabarti, R. Bruce, D. Kershaw, A. Reid, M. Wilder, and K. Flautner, "From SODA to Scotch: The evolution of a wireless baseband processor," in *Proceedings of the 41st Annual IEEE/ACM International Symposium on Microarchitecture*, MICRO 41, (Washington, DC, USA), pp. 152–163, IEEE Computer Society, 2008.
[4] M. Boettcher, B. M. Al-Hashimi, M. Eyole, G. Gabrielli, and A. Reid, "Advanced SIMD: Extending the reach of contemporary SIMD architectures," in *Proceedings of the Conference on Design, Automation & Test in Europe*, DATE '14, (3001 Leuven, Belgium, Belgium), pp. 24:1–24:4, European Design and Automation Association, 2014.
[5] R. M. Russell, "The CRAY-1 computer system," *Commun. ACM*, vol. 21, pp. 63–72, Jan. 1978.
[6] S. S. Baghsorkhi, N. Vasudevan, and Y. Wu, "Flexvec: auto-vectorization for irregular loops," in *Proceedings of the 37th ACM SIGPLAN Conference on Programming Language Design and Implementation*, pp. 697–710, ACM, 2016.
[7] L. Seiler, D. Carmean, E. Sprangle, T. Forsyth, M. Abrash, P. Dubey, S. Junkins, A. Lake, J. Sugerman, R. Cavin, *et al.*, "Larrabee: a many-core x86 architecture for visual computing," in *ACM Transactions on Graphics (TOG)*, vol. 27, p. 18, ACM, 2008.
[8] A. Sodani, "Knights landing (knl): 2nd generation intel® xeon phi processor," in *Hot Chips 27 Symposium (HCS), 2015 IEEE*, pp. 1–24, IEEE, 2015.
[9] "A sneak peak into SVE and VLA programming." https://developer.arm.com/hpc/a-sneak-peek-into-sve-and-vla-programming.
[10] C. Lattner and V. Adve, "The llvm compiler framework and infrastructure tutorial," in *International Workshop on Languages and Compilers for Parallel Computing*, pp. 15–16, Springer, 2004.
[11] https://github.com/benchmark-subsetting/NPB3.0-omp-C.
[12] http://crd.lbl.gov/departments/computer-science/PAR/research/previous-projects/torch-testbed.
[13] https://asc.llnl.gov/computing_resources/purple/archive/benchmarks.
[14] https://asc.llnl.gov/sequoia/benchmarks.
[15] http://accc.riken.jp/en/supercom/himenobmt.
[16] https://bitbucket.org/hpgmg/hpgmg.
[17] https://mantevo.org.
[18] https://asc.llnl.gov/CORAL-benchmarks.